\documentclass[a4paper]{article}

\usepackage{xfrac}
\usepackage{CJKutf8}
\usepackage{ISCSLP2021}
\usepackage{amsmath}
\usepackage{multirow}
\usepackage{cleveref}

\title{Audio Caption in a Car Setting with a Sentence-Level Loss}
\name{Xuenan Xu, Heinrich Dinkel, Mengyue Wu, Kai Yu\thanks{Mengyue Wu and Kai Yu are the corresponding authors. This work has been supported by the Major Program of National Social Science Foundation of China (No.18ZDA293). Experiments have been carried out on the PI supercomputer at Shanghai Jiao Tong University.}}

\address{MoE Key Lab of Artificial Intelligence\\
SpeechLab, Department of Computer Science and Engineering\\ 
Shanghai Jiao Tong University, Shanghai, China}
\email{\{wsntxxn, richman, mengyuewu, kai.yu\}@sjtu.edu.cn}

\begin{document}

\maketitle
\begin{abstract}
Captioning has attracted much attention in image and video understanding while a small amount of work examines audio captioning. 
This paper contributes a Mandarin-annotated dataset for audio captioning within a car scene. 
A sentence-level loss is proposed to be used in tandem with a GRU encoder-decoder model to generate captions with higher semantic similarity to human annotations.
We evaluate the model on the newly-proposed \textit{Car} dataset, a previously published Mandarin \textit{Hospital} dataset and the \textit{Joint} dataset, indicating its generalization capability across different scenes. 
An improvement in all metrics can be observed, including classical natural language generation (NLG) metrics, sentence richness and human evaluation ratings.
However, though detailed audio captions can now be automatically generated, human annotations still outperform model captions on many aspects.
\end{abstract}


\section{Introduction}

Automatic captioning is a challenging task that involves joint learning of different modalities. 
For example, image captioning requires extracting features from an image and combining them with a language model to generate reasonable sentences to describe the image. 
Similarly, video captioning learns features from a temporal sequence of images as well as audio to generate captions.
However, since audio captioning is a relatively new field, it does not attract much attention like image- and video captioning.

\begin{figure}
    \centering
    \includegraphics{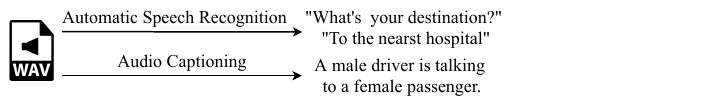}
    \caption{Illustration of the difference between goals of ASR and audio captioning.}
    \label{fig:ASR_audio_caption}
\end{figure}

One well-known task within audio processing, which is commonly associated with audio captioning, is Automatic Speech Recognition (ASR).
There are two main characteristics of audio captioning compared with ASR: 1) audio captioning focuses on all sound events in an audio while ASR only focuses on speech (speech does not necessarily appear in the input to an audio captioning model) 2) audio captioning is an automatic summarization of the audio sound events while ASR directly outputs transcriptions of human speech in the audio.
A comparison of the two tasks' goals is shown in \Cref{fig:ASR_audio_caption}.

Though the success of an audio captioning task in the recent DCASE2020 challenge has prompted a plethora of novel approaches and papers~\cite{koizumi2020_t1,wuyusong2020_t6,wang2020_t6,xu2020_t6}, limited attention is paid to audio captioning within Chinese language processing.  
A Mandarin-annotated 10 hour audio dataset within a hospital scene in conjunction with a baseline encoder-decoder model to generate natural language captions has recently been published~\cite{wu2019audio}.
Although the model performance evaluated by BLEU score is particularly high, human evaluation tells a different story. 
Most machine-generated Mandarin captions are monotonous and repetitive, while by contrast human annotations are much more specific in content and vivid in expression. 
Therefore, a captioning model should endeavor to generate various sentences that not only describe detailed audio content but also contain richer vocabulary and diverse sentence structures. 
For example, for a sound of a car crash in an audio clip, a well-performing model is expected to generate a caption like ``The car went into a crash with others'' or "A traffic accident happened", instead of repetitive ``There is a sound of a car crash" or even ``There are car sounds''.

To achieve the goal of generating specific captions with various expressions, we first publish a dataset on car scene with five annotations per audio. 
Followed by that, we address the variety lacking problem by incorporating an additional sentence-level loss during training.
Similar sequence-level loss has been proved effective in previous work~\cite{prabhavalkar2018minimum,ranzato2015sequence}.
The sentence-level loss is based on context-aware sentence embeddings of diverse, vivid human annotations.
Since there is a supervision signal from the overall sentence-level similarity with human annotations, the model is expected to generate more diverse sentences with similar meaning.
In addition to classical natural language generation (NLG) metrics, we also evaluate our model by output richness, represented by the ratio between the number of unique sentences  and total predicted sentences.
Human evaluation is further performed to subjectively rate the output quality.
Finally, with the newly proposed dataset on car scene, we evaluate our model's generalization capabilities.


\section{Related Work}

\begin{figure*}[t]
    \centerline{\includegraphics[width=\textwidth]{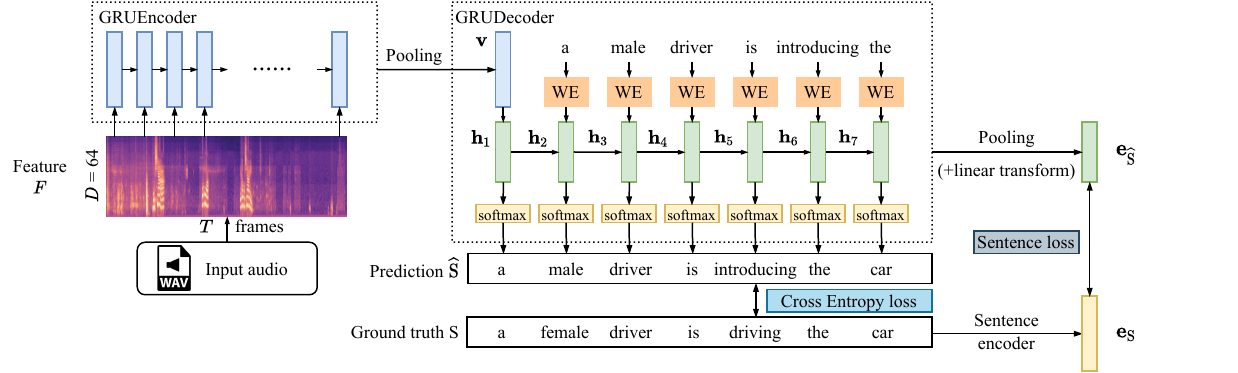}}
    \caption{Our proposed encoder-decoder model with sentence-level loss. Both the encoder and the decoder are single-layer GRUs. The encoder outputs a fixed-sized audio embedding $\mathbf{v}$ from the $T \times D$ input feature $\mathbf{F}$. Then the decoder predicts the sentence with $\mathbf{v}$ as the input to the first timestep. In addition to the standard cross entropy loss between the prediction $\hat{\text{S}}$ and the ground truth $S$, we propose a sentence-level loss, which is the dissimilarity between the prediction embedding $\mathbf{e}_{\hat{\text{S}}}$ and the ground truth embedding $\mathbf{e}_\text{S}$ .} 
    \label{fig:model}
\end{figure*}

\textbf{Audio Captioning Datasets}
To date, a few datasets including Audiocaps~\cite{kim-etal-2019-audiocaps} and Clotho~\cite{drossos:icassp_b:2020} for audio captioning have been published. Most of the current audio captioning datasets are in English. The only existing Mandarin-annotated dataset is the previously mentioned \textit{Hospital Scene} dataset~\cite{wu2019audio}, a detailed comparison with the proposed \textit{Car Scene} dataset will be provided in \Cref{sec:datacar}.

\textbf{Captioning Model}  
Image and video captioning have witnessed promising improvements recently. 
The development of sequence-to-sequence models enables well-performing video captioning models by simply using temporal image information~\cite{pasunuru2017multi}. Later, the attention mechanism is utilized to fuse audio with video information and assign different importance to time frames~\cite{hori2017attention,hori2018multimodal,chuang2017seeing}. 
Shen \emph{et al.}~\cite{shen2017weakly} generates multiple captions in different detail levels and temporal attention.

\textbf{Sentence Embedding}
Early works like GloVe~\cite{pennington2014glove} and Word2Vec~\cite{le2014distributed} in natural language processing (NLP) focus on context-free embedding of words. 
Recently, models like Cove~\cite{mccann2017learned}, ELMo~\cite{peters2018deep} and GPT~\cite{radford2018improving} make use of the self-attention mechanism and transformers to build context-sensitive word representations. 
An unsupervised, C-BOW-like method to embed sentence to fixed-length vector~\cite{pagliardini2018unsupervised} is later proposed.
In this paper, our work is based on the state-of-art sentence embedding technique from BERT~\cite{devlin2019bert}. 
It contains large bidirectional transformers trained on a huge corpus, thus embeddings extracted from the pretrained BERT model perform well in many tasks.

\textbf{Evaluation Metrics} In previous captioning work, evaluation metrics are mainly borrowed from NLG tasks like machine translation and summarization: BLEU@1-4, METEOR, CIDEr and ROUGE-L scores. 
Most of these metrics are based on N-Gram overlaps between model predictions and human references~\cite{vinyals2015show,xu2015show,hori2017attention,chuang2017seeing}. 
In addition, more effective metrics have also been explored. 
\cite{karpathy2015deep} and~\cite{hodosh2013framing} treat image captioning as a sentence ranking task and use recall@k and median@r as their metric. 
Chuang \emph{et al.}~\cite{chuang2017seeing} embeds sentences to fixed length vectors, based on which, a cosine similarity between model predictions and human annotations is involved as a semantic evaluation. 
Our sentence-level loss function is thus inspired to focus on semantic similarity rather than individual wording choices. 

\section{An audio caption dataset in a car scene}
\label{sec:datacar}

This work publishes a 10 hours' Mandarin-annotated dataset on car scene that enables audio captioning. 
English translations using Baidu translator are also provided for broader accessibility. 
The proposed Car dataset contains 3602 car-scene related audio clips, each lasting for 10s.
Each audio clip is annotated by five native Mandarin speakers with a concise labeling method: only natural sentence annotations are included while other metadata are generated from the annotations, e.g., sound events, subjects, etc.

\begin{table}[htpb]
    \centering
    \caption{A comparison of existing audio captioning datasets.}
    \begin{tabular}{|c|c|c|c|c|}
    \hline
      Dataset & Language & Scene & \# Audios & \# Captions \\
      \hline
      AudioCaps & English & General & 39,597 & 45,513 \\
      Clotho & English & General & 4,981 & 24,905 \\
      Hospital & Mandarin & Specific & 3,709 & 11,121 \\
      Car & Mandarin & Specific & 3,602 & 18,010 \\
    \hline 
    \end{tabular}
    \label{tab:datacompare}
\end{table}

This dataset exhibits a handful of discrepancies from the previously published datasets (see \Cref{tab:datacompare}): 1) We provide scene-specific datasets for precise caption generation, in comparison with general purpose datasets like AudioCaps~\cite{kim-etal-2019-audiocaps} and Clotho~\cite{drossos:icassp_b:2020}; 2) The proposed Car dataset includes large quantities of real-life recordings which are suitable for real applications while the \textit{Hospital} dataset~\cite{wu2019audio} consists of more video clips from TV shows due to limited surveillance access in hospital.
  
    \begin{table}[htpb]
        \centering
        \caption{Most Frequent Sound Events.}
        \label{tab:tag_analysis}
        \begin{tabular}{|c|c|c|}
            \hline
           Rank  &  Sound Event  & \# of events \\\hline
           1  &  Engine Sound  &  1442 \\
           2  &  Noise       &  872 \\
           3  &  Clicking Sound &  812 \\
           4  &  Music        &  798 \\
           5  &  Speech        &  563 \\
           \hline
        \end{tabular}

    \end{table}

\Cref{tab:tag_analysis} shows the top 5 sound events of the proposed Car dataset, indicating that the sound events are quite scene-specific.
The Car dataset is split into a development set and an evaluation set, which encompasses 3241 and 361 audio clips respectively. 
High sentence diversity is observed in both sets: 
only 6.7\% annotations in the development set and 1.9\% in the evaluation set are repeated.
From the distribution of the top 5 tokens in \Cref{tab:token_distribution} it can be seen that the development-evaluation split exhibits a similar token distribution.

\begin{CJK}{UTF8}{gbsn}
\begin{table}[htpb]
    \centering
    \caption{Token Distribution in the proposed Car dataset.}
    \begin{tabular}{|c|c|c|c|}
    \hline
        Rank    &   Token   &   Dev \%   &   Eval \% \\ 
       \hline
      1 &   is/are 在  &  6.01   & 6.01  \\ 
      2 &   driving 行驶  &  5.37     & 5.55 \\ 
      3 &   automobile 汽车  &  5.01   & 5.11 \\
      4 & 's 的         &   4.01    & 4.58  \\
      5 &   driver 司机 &  3.35 & 3.45 \\
      \hline\hline
      \multicolumn{2}{|c|}{mean \# of tokens} &  14.21 & 14.03 \\
      \hline
    \end{tabular}
    \label{tab:token_distribution}
\end{table}
\end{CJK}

\section{Model Description}
Since the previously utilized GRU encoder-decoder model~\cite{wu2019audio} can generate audio relevant and grammatically correct sentences, we continue to incorporate a similar architecture with certain modifications for further performance enhancement. 
Our architecture consists of an audio embedder (encoder) and a text generator (decoder).
\paragraph*{Encoder} For each audio clip, our GRU encoder reads a log mel spectrogram (LMS) feature $F$ and encodes it into a fixed-length feature vector $\mathbf{v}$. 
\paragraph*{Decoder} The decoder takes the audio embedding $\mathbf{v}$ as its input to generate natural language captions.
However, our decoder network works differently during training and evaluation. 
During training, when annotated captions are available, teacher forcing is used to accelerate the training process. 
$\mathbf{v}$ is thus concatenated with the ground truth annotation embeddings as the input to the decoder.
Without annotation access during evaluation, $\mathbf{v}$ is directly fed to the decoder.
For every timestep, the decoder generates a single token until the ``$<$EOS$>$'' (end of sentence) token is generated (see \Cref{fig:model}).

In this paper, we utilize two loss functions during training: 1) standard Cross Entropy (CE) loss 2) the newly proposed sentence-level loss. 
Using a word and sentence loss combination, our model is expected to not only focus on wording selection but also sentence-level semantic similarity, which eventually leads to captions with human-like content while being diversified in sentence structure.

\textbf{CE loss}
Standard cross entropy is used as the word-level loss \Cref{eq:CE_loss}, which is defined as the negative log likelihood of the expected word $\text{S}_t$ given the input audio feature $F$ and the model parameters $\theta$ at time $t$.  

\begin{equation}
    \ell_{\text{CE}}(\theta;\text{S},F) = -\sum_{t=1}^T \log p(\text{S}_t|\theta,F)
    \label{eq:CE_loss}
\end{equation}

\textbf{Sentence-level loss}
In addition to the standard CE loss at word-level, we propose a novel sentence-level loss to capture semantic similarity better. 
Since the decoder outputs a hidden state ($\mathbf{h}_t$) at each timestep $t$, we first pool the hidden states of all timesteps to get a single representation of the prediction. 
As \Cref{eq:mean_pooling} shows, we use mean pooling on all $\mathbf{h}_t$ to obtain the representation $\mathbf{e}_{\hat{\text{S}}}$.

\begin{equation}
    \label{eq:mean_pooling}
    \mathbf{e}_{\hat{\text{S}}}(\theta,F) = \frac{1}{T}\sum_{t=1}^T \mathbf{h}_t(\theta,F)
\end{equation}

In order to minimize the embedding difference between $\mathbf{e}_{\hat{\text{S}}}$ and annotated sentences ($\mathbf{e}_\text{S}$), we develop a sentence loss function opposed to cosine similarity (see \Cref{eq:sentence_loss}, where $\epsilon$ is a small number ensuring numerical stability). 
In this way, a small sentence loss indicates a high semantic similarity. 
In cases where $|\mathbf{e}_{\hat{\text{S}}}|$ differs from $|\mathbf{e}_\text{S}|$, a linear transformation layer is added after the mean pooling operation to ensure $\mathbf{e}_{\hat{\text{S}}}$ and $\mathbf{e}_\text{S}$ are of equal dimension. 

\begin{equation}
    \label{eq:sentence_loss}
    \ell_\text{sentence}(\theta;\mathbf{e}_\text{S},F) = 1 - \dfrac{\mathbf{e}_\text{S} \cdot \mathbf{e}_{\hat{\text{S}}}(\theta,F)}{\max(\Vert \mathbf{e}_\text{S} \Vert _2 \cdot \Vert \mathbf{e}_{\hat{\text{S}}}(\theta,F) \Vert _2, \epsilon)}
\end{equation}

Accordingly, the training objective (\Cref{eq:combined_loss}) minimizes the weighted sum of the word (\Cref{eq:CE_loss}) and sentence (\Cref{eq:sentence_loss}) loss, where $\alpha$ is a fixed hyperparameter.

\begin{equation}
    \label{eq:combined_loss}
    \ell_{\text{combined}}(\theta;\text{S},\mathbf{e}_\text{S},F) = \ell_{\text{CE}}(\theta;\text{S},F) + \alpha \cdot \ell_\text{sentence}(\theta;\mathbf{e}_\text{S},F)
\end{equation}

\section{Experiments}
\label{sec:experiments}
\subsection{Datasets}
We first validate our proposed \textit{Car} dataset for its effectiveness in audio captioning. 
To investigate the generalization capabilities of our model and the proposed sentence-level loss, in particular under cross-scene circumstances, we further experiment on another two datasets. 
One is the \textit{Hospital} dataset, including 3709 audio clips with three human annotations for each audio clip; the other is the creation of a \textit{Joint} dataset that merges the \textit{Car} and \textit{Hospital} datasets.
It should be noted that the \textit{Joint} dataset is domain balanced since the number of audio clips within the two datasets are similar (Car: 3602; Hospital: 3709).

\subsection{Data preprocessing}
Standard 64 dimensional LMS features from a 40 ms window are extracted every 20 ms.
During training we apply global standardization (mean and variance) on each feature.
Since the annotations are in Mandarin Chinese, a language that does not separate words by space in sentences, the annotations need to be tokenized.
Here, Stanford core NLP tools~\cite{qi2018universal} are used for parsing. 
We also use the public simplified and traditional Chinese BERT model\footnote{https://storage.googleapis.com/bert\_models/2018\_11\_03/chinese\_L-12\_H-768\_A-12.zip} to obtain the fixed-length annotation embedding $\mathbf{e}_\text{S}$ for sentence-level loss training.

\subsection{Training Details}

Both the encoder and the decoder are composed of a single-layer GRU with a hidden size of 512.
The dimension of $\mathbf{v}$ is 256.
BERT encodes S into a 768 dimensional $\mathbf{e}_{\text{S}}$.
The development set is further split into a training subset and a validation subset with a ratio of $9:1$.
Training is done using the Adam optimization algorithm~\cite{Kingma2014} with an initial learning rate $4\mathrm{e}{-4}$, batch size of 32 and default beta values given by the pytorch framework~\cite{paszke2019_nips}.
$\alpha$ is set to 10 in combined loss training and the model is trained for a fixed amount of 25 epochs.
Following previous work on image caption~\cite{rennie2017self}, we calculate the CIDEr score on the validation set after each epoch and choose the model with the highest CIDEr score for evaluation.

\subsection{Results}

Results are analyzed from two aspects: 1) the model performance evaluated by different metrics; 2) the model generalization capabilities on different datasets.

\subsubsection{Evaluation Metrics}

The presentation of our results is split into 1) Objective metrics, including BLEU@1-4, ROUGE and CIDEr; 2) Human Evaluation, involving eight native speakers' ratings on machine- and human-generated captions. 

\paragraph*{Objective Metrics} 

\begin{table}[t]
    \centering
    \caption{Results on the Car evaluation set. $\ell_\text{CE}$ and $\ell_\text{combined}$ correspond to different loss functions (see \Cref{eq:CE_loss} and \Cref{eq:combined_loss}).}
    \begin{tabular}{|c|c|c|}
    \hline
      \multirow{2}{*}{Metric}       & \multicolumn{2}{c|}{Training loss}\\ 
      \cline{2-3}
       & $\ell_\text{CE}$ & $\ell_\text{combined}$\\
      \hline 
      $\text{BLEU}_1$   & \textbf{0.720} & 0.706\\  
      $\text{BLEU}_2$   & 0.549 & \textbf{0.553}\\ 
      $\text{BLEU}_3$   & 0.415 & \textbf{0.433}\\ 
      $\text{BLEU}_4$   & 0.322 & \textbf{0.348}\\ 
      $\text{ROUGE}_\text{L}$ & 0.491  & \textbf{0.518}\\ 
      CIDEr   & 0.364  & \textbf{0.447}\\ 
      Richness & 0.116 & \textbf{0.213} \\
      \hline
    \end{tabular}
    \label{results}
\end{table}

Classical NLG evaluation metrics include BLEU@1-4, $\text{ROUGE}_\text{L}$, CIDEr, METEOR and SPICE.
However, METEOR and SPICE depend on a paraphrasing library named WordNet while there is no such library for Mandarin.
Therefore, we only present BLEU@1-4, $\text{ROUGE}_\text{L}$ and CIDEr here.
In addition to these scores, we also directly count the unique caption number in all predictions, indicating the output richness.
\Cref{results} illustrates the objective results.
Except $\text{BLEU}_1$, the model trained with $\ell_{\text{combined}}$ achieves a performance improvement on all evaluation metrics compared with $\ell_{\text{CE}}$ trained model.
The most significant improvement lies in CIDEr, achieving a 22.8\% relative gain.
The improvement in objective metrics indicates that sentence-level loss is helpful in training the model to output content-correlated sentences. 
In addition, richness of the model predictions also increases from 0.116 to 0.213.
Sentences generated by $\ell_{\text{combined}}$ trained model may have similar semantic meanings but are diverse in expression. \\
An example  is provided here:

\indent\textbf{Human Annotation:} When the car is moving, a man is talking with music on the car and the sound of braking.\\
\indent\textbf{CE Loss Prediction:} When the car is moving, the male driver is talking with the female passenger accompanied by the engine sound.\\
\indent\textbf{Combined Loss Prediction:} The car is moving with music playing. The male driver is talking and suddenly the car hits another one.\\

\paragraph*{Human Evaluation}
Eight native Mandarin speakers are invited to evaluate the model predictions. 
We randomly pick five audios for evaluation.
Human annotations, $\ell_\text{CE}$ predictions and $\ell_\text{combined}$ predictions are evaluated. 
Raters score each caption on a five-point scale, where 1 stands for the least and 5 signifies the most useful. 
Results show that human annotations averaged 4.05, followed by $\ell_\text{combined}$ (scored 3.63), with $\ell_\text{CE}$ predictions being the least useful (scored 3.18). 
Although $\ell_\text{combined}$ predictions show significant advantage against $\ell_\text{CE}$ predictions in terms of human evaluation scores, there is still a gap between our model predictions and human annotations. 
Examples are provided at the end of this section.

\subsubsection{Generalization}
In order to verify the generalization capabilities of our model and the combined loss function, we train the model on the other two datasets: \textit{Hospital} dataset and \textit{Joint} dataset. 
Results evaluated by different metrics can be seen in  \Cref{tab:generalization}.
The advantage in description accuracy and diversity are verified.
Firstly, the GRU encoder-decoder model with our proposed sentence-level loss can be generalized to other datasets.
There is an improvement on all metrics for both datasets, comparing $\ell_\text{combined}$ with the $\ell_\text{CE}$ baseline.
Specifically, for the \textit{Joint} dataset, annotations on different scenes are mixed while the improvement is still significant, indicating that the proposed sentence-level loss is not only effective on a specific scene. 
Secondly, the current model is capable of generating richer sentences. 
On both datasets, there is an about 30\% relative increase in richness compared with the baseline model trained with only CE loss. 

\begin{table}[htpb]
    \centering
    \caption{Results on Hospital and Joint datasets, trained by different loss functions.}
    \begin{tabular}{|c|c|c|c|c|c|}
    \hline
    \multicolumn{2}{|c|}{Datasets} & \multicolumn{2}{|c|}{Hospital}  & \multicolumn{2}{|c|}{Joint} \\
    \hline
    \multicolumn{2}{|c|}{Training loss} & $\ell_\text{CE}$ & $\ell_\text{combined}$ & $\ell_\text{CE}$ & $\ell_\text{combined}$ \\
    \hline
    \multirow{7}{*}{Metric} & $\text{BLEU}_\text{1}$ & 0.526 & \textbf{0.543} & 0.614 & 0.614\\
    & $\text{BLEU}_\text{2}$ & 0.430 & \textbf{0.432} & 0.235 & \textbf{0.243}\\
    & $\text{BLEU}_\text{3}$ & 0.205 & \textbf{0.229} & 0.311 & \textbf{0.317}\\
    & $\text{BLEU}_\text{4}$ & 0.144 & \textbf{0.166} & 0.235 & \textbf{0.243}\\
    & $\text{ROUGE}_\text{L}$ & 0.389 & \textbf{0.392} & 0.429 & \textbf{0.442}\\
    & CIDEr   &   0.326  & \textbf{0.366} & 0.435  & \textbf{0.512}  \\ 
    & Richness  & 0.429 & \textbf{0.566} & 0.347 & \textbf{0.437} \\
    \hline
    \end{tabular}
    \label{tab:generalization}
   
\end{table}

\begin{CJK}{UTF8}{gbsn}
\noindent\fbox{
\scriptsize{
\parbox{\columnwidth}{
\textbf{\textit{Hyp Score 5: Accurate, comprehensive and vivid description}}\\
\textbf{Hyp:} 汽车在行驶中男司机在和女乘客聊天伴随着发动机声\\
The male driver is chatting with the female passenger while the car is moving.\\
\textbf{Ref 1:} 行车过程中司机和后排乘客说话 \\
The driver and the passenger on the back are talking during driving. (Score 4)\\
\textbf{Ref 2:} 车在行驶中男司机找女乘客搭讪女乘客小声应答  \\
The driver strikes up a conversation with a female passenger while the car is moving. (Score 5)\\

\textbf{\textit{Hyp Score 3: Generally correct, with some missing or redundant description}}\\ 
\textbf{Hyp:} 汽车停在路边男司机在介绍汽车\\
The car parks at the roadside and the male driver is introducing the car.\\
\textbf{Ref 1:} 汽车停靠在马路边司机讲解汽车性能有风噪声 \\
The car parks at the roadside. The driver introduces the car performance along with wind noise. (Score 4)\\
\textbf{Ref 2:} 汽车停在路边男司机对女乘客讲解相关内容有车噪声 \\
The car parks at the roadside. The male driver introduces it to the female passengers along with car noise. (Score 5)\\ 

\textbf{\textit{Hyp Score 1 Not suitable at all}}\\
\textbf{Hyp:} 汽车在行驶中男司机和女乘客在聊天\\
The male driver and the female passenger are chatting while the car is driving\\
\textbf{Ref 1:} 车辆在高速行驶车里在放音乐  \\
The car is running fast with music playing. (Score 3)\\
\textbf{Ref 2:} 汽车行驶中车内放着音乐外面传来物体落在车上的声音汽车停住了 \\
When the car is running with music in it, there is sound outside the car. Then the car stops. (Score 5)\\}
}
}
\end{CJK}

\section{Conclusion}

In this paper, we propose a 10 hour long Car scene corpus. 
Further, a sentence-level loss to provide a supervision signal from the sentence semantics is proposed.
Metrics including classical NLG metrics and output richness show that our approach now generates more content-related captions with higher diversity.
Human evaluation results also validate the advantage of the added sentence-level loss.
Validation of the proposed approach is done on three Mandarin audiocaption datasets (Hospital, Car, Joint), verifying its generalization capability.
Despite the effectiveness of our proposed GRU encoder-decoder model with a sentence-level loss, there is still a significant gap between model predictions and human annotations.

\bibliographystyle{IEEEtran}

\bibliography{refs}

\end{document}